\documentclass[pra,twocolumn,preprintnumbers,amsmath,amssymb,superscriptaddress,showpacs,longbibliography]{revtex4-2}
\usepackage[T1]{fontenc} 
\usepackage[utf8]{inputenc} 

\usepackage{graphicx}
\usepackage{latexsym}
\usepackage{amsmath}
\usepackage{graphics}
\usepackage{amssymb}
\usepackage{layout}
\usepackage{verbatim}
\usepackage{amsfonts,epsfig,dsfont}
\usepackage{soul}
\usepackage[dvipsnames,table]{xcolor} 
\usepackage{hyperref}
\usepackage{tikz}
\usepackage{tikz-feynman}
\usepackage{color}
\hypersetup{colorlinks, linkcolor={blue!90!black}, citecolor={blue!90!black}, urlcolor={blue!90!black} }

\newcommand{\beq}{\begin{equation}}
\newcommand{\eeq}{\end{equation}}
\newcommand{\bea}{\begin{eqnarray}}
\newcommand{\eea}{\end{eqnarray}}

\newcommand{\unit}{\mathbb{I}}

\usepackage{color}

\usepackage{ulem} 

\begin{document}

\title{Detection of entanglement during pure dephasing evolutions for systems and environments 
	of any size
}
\author{Ma\l gorzata Strza\l ka}
\author{Katarzyna Roszak}
\affiliation{Department of Theoretical Physics, Faculty of Fundamental Problems of Technology, Wroc{\l}aw University of Science and Technology,
50-370 Wroc{\l}aw, Poland}

\date{\today}

\begin{abstract}
W generalize the scheme for detection of qubit-environment entanglement to qudit-environment systems.
This is of relevance for many-qubit systems and the quantification of the operation of quantum algorithms under the influence of external noise, since only decoherence that is not entangling in its nature 
can be effectively described by quantum channels and similar methods in more complicated scenarios. 
The generalization involves 
an increase of the class of entangled states which are not detected by the scheme, but 
the type of entanglement which cannot be detected 
is also least likely to qualitatively influence decoherence.
We exemplify the operation of the scheme on a realistically modelled NV-center spin qutrit 
interacting with an environment of nuclear spins.
\end{abstract}
\maketitle

\section{Introduction}

Qubits are the simplest quantum systems, since their Hilbert space contains only two states.
The result of this is that some complex measures
of quantumness or quantum correlations are much easier to study for qubits than for
larger quantum systems. One example of this is mixed state entanglement \cite{nielsen00,plenio07}
which can be found directly from the density matrix for a system of two qubits \cite{wootters98}, but otherwise requires minimization over all possible preparations
of a state \cite{plenio07,horodecki09} or the use of measures which do not quantify all types of entanglement \cite{vidal02,lee00a,plenio05b,nakano13}.
Similarly bound entanglement \cite{horodecki97b,horodecki98}, a type of entanglement which is not detected by the Peres-Horodecki criterion \cite{peres96a,horodecki96}, does not exist for systems of two qubits. 

The number of coherences
(off-diagonal elements of the density matrix)
grows quadratically with the size of the system (as $N(N-1)/2$ to be precise; obviously a density matrix is Hermitian, hence only half of the coherences are independent
variables), so the single qubit coherence 
is replaced by three coherences for a qutrit, six coherences for a system of size $N=4$, and so on.
Furthermore, the dependencies between the different coherences are relevant. An example here
is the simple task of checking if a matrix is a density matrix and can therefore describe
a physical state. This requires checking three conditions: Hermitianity, unit trace, and positivity.
Only the third conditions is problematic as it requires diagonalization of the matrix, which can only
be done numerically for larger matrices. For a two by two Hermitian matrix only the absolute value of 
the coherence is relevant for positivity, which is not the case already for a qutrit.

The consequence is that there is a qualitative difference when studying larger systems as opposed
to studies restricted to qubits and conclusions drawn for qubits rarely translate seamlessly to 
larger systems. This is also the case for asymmetric bipartite systems
composed of a qudit ($N$ dimensional system of interest) and its environment. 
In particular, entanglement formed between a qudit and its environment is much harder to study
than in the case of a qubit. This is evident for pure-dephasing interactions, the only 
type of system-environment couplings for which simple, general formulas for qualification of
a state obtained during the evolution as entangled or not exist \cite{roszak15,roszak18}.
Qualifying qubit-environment entanglement (QEE) requires checking a single condition \cite{roszak15}
and this allowed an entanglement measure tailored specifically to quantify this type of entanglement
to be proposed \cite{roszak20}, which yields substantial computational advantage 
with respect to standard entanglement measures \cite{plenio07,horodecki09,strzalka20}.
Qualifying system-environment entanglement (SEE) on the other hand requires checking $N-1$
conditions which are analogous to the QEE conditions and additionally $(N-1)(N-2)/2$
conditions which are qualitatively different \cite{roszak18}. This rapid growth in complexity 
with system size $N$ precludes the possibility of an analogous SEE measure to be proposed.

The creation of entanglement with the environment
throughout the evolution is relevant because the behavior of the environment is qualitatively
different when entanglement is formed and when the evolution is separable \cite{roszak15,roszak17,roszak18}.
In many situations QEE can lead to 
effects which cannot be explained by decoherence modeled classically \cite{roszak15b}.
This backaction, the situation when entanglement manifests itself in the state of the 
environment, which in turn influences the evolution of the qubit, is the reason why QEE
can be measured with little effort \cite{roszak19a,rzepkowski20}.

In the following we will study a qudit for which the interaction with the environment leads
to pure dephasing, in order to generalize a scheme for the detection of QEE by operations
and measurements only on the qubit \cite{roszak19a}. 
This type of interaction is the dominating decoherence mechanism
for many solid state qubits
\cite{medford12,kawakami16,zhao12,kwiatkowski18,malinowski17,knee16,yulin18,touzard19}.
One motivation for the importance of SEE is that most solid state qubits are in fact only approximations of qubits 
(e.~g.~where two states are energetically distinct and can therefore be addressed separately).
The more relevant one is that ensembles of qubits are of vital importance for any type of quantum
data processing and ensembles of qubits interacting with an environment can no longer be 
treated with the methods for studying QEE. From the point of view of entanglement with an environment
they are in fact qudits and display the whole range of complexity relating to many coherences
and phase relations between them.

We will show that one method for the detection of QEE \cite{roszak19a} can in fact be generalized 
to detect SEE. The complexity of the procedure only grows linearly with the size of the qudit, so
it does not reflect the quadratic growth of the number of SEE criteria \cite{roszak18}.
The price to pay is the growing number of entangled states that cannot be detected by the procedure.
Additionally to the type of entanglement which cannot be detected by the qubit procedure,
there is now a second class of entanglement which cannot be witnessed for larger systems.
Optimistically, the type of entanglement which is detected by the procedure is the type
which is most likely to influence the operation of quantum algorithms \cite{rzepkowski20}.

The paper is organized as follows.
In Sec.~\ref{sec2} we introduce the type of system-environment density matrix which can be 
classified in terms of SEE by the proposed scheme and the conditions on the Hamiltonian and 
initial state of the system and the environment to guarantee this form throughout the evolution.
In Sec.~\ref{sec3} we restate criteria for separability of such density matrices. We introduce
the proposed scheme for the detection of entanglement in Sec.~\ref{sec4} and study the limitations
of applicability of the scheme in Sec.~\ref{sec5}. In Sec.~\ref{sec6} we study the working of
the scheme on an NV-center spin qutrit interacting with a nuclear environment. Sec.~\ref{sec7} concludes
the paper.

\section{Class of problems studied \label{sec2}}
In the following we will present a scheme which allows to detect entanglement between
a quantum system of interest (with no limitation on the dimension of its Hilbert space)
and its environment.
The method can only be used for system-environment density matrices
of the form ($N$ is the dimension of the system, the dimension of the environment is unspecified
and arbitrary)
\begin{equation}
\label{sigma}
\hat{\sigma}
=\sum_{k,l=0}^{N-1}c_kc^*_l
|k\rangle\langle l|\otimes\hat{R}_{kl}.
\end{equation}
Here the states on the left side of the tensor product correspond to some basis $\{|k\rangle\}$
in the system subspace, while the matrices $\hat{R}_{kl}$ describe the environment. For the full
matrix (\ref{sigma}) to be a density matrix, the diagonal environmental matrices $\hat{R}_{kk}$
have to be density matrices, but there is no such limitation for off-diagonal matrices,
with $k\neq l$.

Although density matrices of the form (\ref{sigma}) are, at certain time-instants,
encountered in evolutions
governed by different Hamiltonians \cite{mazurek14b},
the prevailing situation when they are encountered is when the system-environment Hamiltonian
can only lead to pure-dephasing decoherence of the system.
A system-environment Hamiltonian of this class can always be written in the form \cite{roszak18}
\begin{equation}
\label{ham0}
\hat{H}=
\sum_{k=0}^{N-1}\varepsilon_k|k\rangle\langle k|+\hat{H}_{\mathrm{E}}+
\sum_{k=0}^{N-1}|k\rangle\langle k|\otimes{\hat{V}_k},
\end{equation}
where $\{|k\rangle\}$ is the same system basis as used in eq.~(\ref{sigma})
and is now specified as the pointer basis of the system \cite{zurek81,zurek03}.
Obviously the first term in the Hamiltonian (\ref{ham0}) is the free Hamiltonian of the system,
the second term is the (arbitrary) free Hamiltonian of the environment, and the third term
describes the evolution. The first and last terms commute, which is the necessary and sufficient
condition for the Hamiltonian to lead to pure dephasing for all initial states
(such Hamiltonians cannot describe processes which involve energy exchange between
the system and the enviornment).

A Hamiltonian of this type is diagonal in the subspace of the system, and the corresponding 
evolution operator retains this property,
\begin{equation}
\label{u}
\hat{U}(t)=\sum_{k=0}^{N-1}|k\rangle\langle k|\otimes{\hat{w}_k(t)}.
\end{equation}
The environmental operators $\hat{w}_k(t)$ can be understood as evolution operators
of the environment conditional on the pointer state of the
system and are given by
\begin{equation}
\label{w0}
\hat{w}_k(t)=e^{-\frac{i}{\hbar}\varepsilon_kt}e^{-\frac{i}{\hbar}(\hat{H}_{\mathrm{E}}+\hat{V}_k)t}.
\end{equation}
The free evolution of each pointer state is included in the operators (\ref{w0}), but it has no
bearing on entanglement and as such is irrelevant for the results presented here.

Using eq.~(\ref{u}) on any initial system-environment state will yield their joint density
matrix at time $t$, but to obtain a density matrix of the form (\ref{sigma}), restriction
on the initial state is needed. Firstly, the state must be of product form with respect to the
environment, and secondly, the initial system state must be pure
\begin{equation}
\label{ini}
\hat{\sigma}(0)=|\psi\rangle\langle\psi|\otimes\hat{R}(0),
\end{equation}
with $|\psi\rangle=\sum_{k=0}^{N-1}c_k
|k\rangle$. The initial state of the environment $\hat{R}(0)$ is arbitrary.
Acting with the evolution operator (\ref{u}) on the initial state (\ref{ini}),
we obtain a system-environment density matrix of the form (\ref{sigma})
for all times $t$, and the environmental matrices, which are the only time-dependent element,
are given by
\begin{equation}
\label{r}
\hat{R}_{kl}(t)=\hat{w}_k(t)\hat{R}(0)\hat{w}_l^{\dagger}(t).
\end{equation}

\section{Criteria for system-environment separability \label{sec3}}

In Ref.~\cite{roszak18} it has been shown that to qualify 
a system-environment state of the form (\ref{sigma}) as entangled or separable
it is enough to check two classes of criteria, which have been derived
from the Peres-Horodecki criterion \cite{peres96a,horodecki96}
and the definition of mixed state separability.

The following are criteria of separability, and if any of them is violated then there 
is entanglement between the system and its enviornment.
The first class of criteria is a generalization of the (single) separability
criterion of QEE \cite{roszak15},
and states that separability requires that for all $k\neq l$ we have
\begin{equation}
\label{war1}
\hat{R}_{kk}(t)=\hat{R}_{ll}(t).
\end{equation}
There are $N-1$ independent criteria of this type \cite{roszak18},
where $N$ is the dimension of the system and it is enough to check (\ref{war1})
with $l$ set constantly to a given value e.~g.~$l=0$. 
Physically, if criterion (\ref{war1}) is fulfilled for a given $k$ and $l$, it means
that the evolution of the environment is indistinguishable regardless if the system is in pointer
state $|k\rangle$ or $|l\rangle$. If all of such criteria are met then the environment evolves
in exactly the same way for the system in any of the pointer states. Contrarily to pure initial
states of the environment, this does not preclude decoherence of the system
which is not initially in a pointer state (or mixture thereof) \cite{eisert02,roszak15,roszak18}. 

The second class of criteria requires commutation between products of different conditional
evolution operators of the environment (\ref{w0}), namely for separability we must have
\begin{equation}
\label{war2}
\left[\hat{w}_i(t)\hat{w}_j^{\dagger}(t),\hat{w}_k(t)\hat{w}_l^{\dagger}(t)\right]=0
\end{equation}
for all $i$, $j$, $k$, and $l$. Only $(N-1)(N-2)/2$ of these conditions are independent \cite{roszak18}.

The second class of separability criteria lacks the straightforward physical interpretation 
characteristic for the first class, which correlates SEE 
with information about the system state that has been transferred into the environment.
This correlation allows for the detection of entanglement at least in principle, by measurements 
performed on the environment. There exist states of the form (\ref{sigma})
for which all of the separability criteria of the first type are fulfilled, but not all of the criteria of the 
second type; such states are entangled \cite{roszak18}.
 
\section{Scheme for detection of SEE \label{sec4}}

For a qubit system, there exists only one separability criterion
and it is of the first type (\ref{war1}). In this case the distinguishability of entangled and separable
states by measurements on the environment alone, can be used to design schemes for entanglement
detection which are operated solely on the qubit \cite{roszak19a,rzepkowski20}. This is a result
of the back-action of the environment on the evolution of the qubit and the possibility of
preparing a state of the environment by allowing it to evolve in the presence of the system in one
of its pointer states.
If the qudit environment state (\ref{sigma}) is entangled in such a way that it violates 
any of the separability criteria (\ref{war1}) then this type of entanglement can also by detected
by operations and measurements restricted to the system.

The procedure for the detection of QEE described in Ref.~\cite{roszak19a}
is particularly straightforward to generalize. To detect if there is entanglement in 
qudit-environment state given by eq.~(\ref{sigma}) at time $\tau$ which is obtained using the evolution operator (\ref{u}) on initial state (\ref{ini}), one must prepare and measure modified qudit-environment states, which involve a preparation of the environment prior to exciting a superposition system state.
The idea is as follows. At time $t=0$ the system is prepared in one of its pointer states 
$|k\rangle$ and allowed to evolve for time $\tau$. This does not change the state of the system
but the environment does evolve, so the system-environment state is given by
\begin{equation}
\hat{\sigma}(\tau)=|k\rangle\langle k|\otimes\hat{R}_{kk}(\tau).
\end{equation}
If the system is now (at time $\tau$) prepared in a superposition state $|\psi\rangle=\sum_{k=0}^{N-1}c_k
|k\rangle$, then it will evolve according to eq.~(\ref{sigma}), but with a new initial state,
$\hat{R}_{kk}(\tau)$ instead of $\hat{R}(0)$.
Further evolution will lead to pure dephasing of the qudit and each of its coherences
will evolve according to
\begin{equation}
\label{rhoij}
\rho_{ij}^{(k)}(\tau,t)=c_ic_j^*\mathrm{Tr}\left(
\hat{w}_i(t)\hat{w}_k(\tau)\hat{R}(0)
\hat{w}_k^{\dagger}(\tau)\hat{w}_j^{\dagger}(t)\right),
\end{equation}
where $t$ is the time elapsed from time $\tau$. 
An ideal test state $|\psi\rangle$ is an equal superposition of all pointer states
as it maximizes the chances of determining entanglement.

If the procedure is repeated for a different initial system pointer state $|l\rangle$
and any of the coherences (\ref{rhoij}) 
show a different evolution at any point after time $\tau$,
$\rho_{ij}^{(k)}(\tau,t)\neq \rho_{ij}^{(l)}(\tau,t)$,
this signifies that at time $\tau$ the 
criterion (\ref{war1}) is not fulfilled for states $|k\rangle$
and $|l\rangle$.
This further means that if the system was initialized in any superposition
which contains pointer states $|k\rangle$ and $|l\rangle$
and the environment was initialized in the state $\hat{R}(0)$,
then at time $\tau$ the joint system-environment state would be entangled.

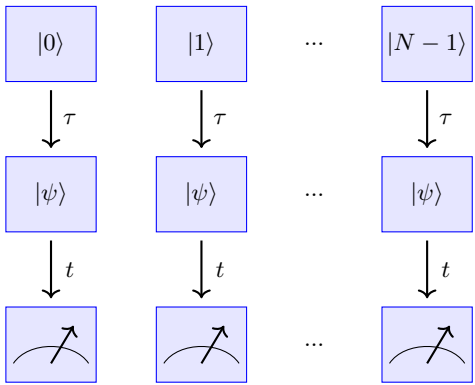
\begin{figure}[t]
\begin{center}
	\begin{tikzpicture}
	\filldraw[fill=blue!10!white, draw=blue!120]  (0.4,2) rectangle (1.6,3);
	\filldraw[fill=blue!10!white, draw=blue!120] (2.4,2) rectangle (3.6,3);
	\filldraw[fill=blue!10!white, draw=blue!120] (5.4,2) rectangle (6.6,3);
	\filldraw[fill=blue!10!white, draw=blue!120]  (0.4,4) rectangle (1.6,5);
	\filldraw[fill=blue!10!white, draw=blue!120] (2.4,4) rectangle (3.6,5);
	\filldraw[fill=blue!10!white, draw=blue!120] (5.4,4) rectangle (6.6,5);
	\node[] (a) at (1,1) {};
	\node[shape=rectangle, ] (a1) at (1.25,1.5) {$t$};
	\node[] (b0) at (1,2.5) {$|\psi\rangle$};
	\node[] (b) at (1,2) {};
	\node[shape=rectangle, ] (a1) at (1.25,3.5) {$\tau$};
	\node[] (c) at (1,3) {};
	\node[] (cc) at (1,4) {};
	\node[] (c0) at (1,4.5) {$|0\rangle$};
	\node[] (d) at (3,1) {};
	\node[shape=rectangle, ] (a1) at (3.25,1.5) {$t$};
	\node[] (e) at (3,2) {};
	\node[] (e0) at (3,2.5) {$|\psi\rangle$};
	\node[shape=rectangle, ] (a1) at (3.25,3.5) {$\tau$};
	\node[] (f0) at (3,4.5) {$|1\rangle$};
	\node[] (f) at (3,3) {};
	\node[] (ff) at (3,4) {};
	\node[] (a1) at (4.5,0.5) {...};
	\node[] (a1) at (4.5,2.5) {...};
	\node[] (a1) at (4.5,4.5) {...};
	\node[] (g) at (6,1) {};
	\node[] (a1) at (6.25,1.5) {$t$};
	\node[] (h0) at (6,2.5) {$|\psi\rangle$};
	\node[] (h) at (6,2) {};
	\node[] (a1) at (6.25,3.5) {$\tau$};
	\node[] (i0) at (6,4.5) {$|N-1\rangle$};
	\node[] (i) at (6,3) {};
	\node[] (ii) at (6,4) {};
	\path[thick,->] (cc) edge (c);
	\path[thick,->] (b) edge (a);
	\path[thick,->] (ii) edge (i);
	\path[thick,->] (h) edge (g);
	\path[thick,->] (ff) edge (f);
	\path[thick,->] (e) edge (d);
	\filldraw[fill=blue!10!white, draw=blue!120]  (0.4,0) rectangle (1.6,1);
	\filldraw[fill=blue!10!white, draw=blue!120] (2.4,0) rectangle (3.6,1);
	\filldraw[fill=blue!10!white, draw=blue!120] (5.4,0) rectangle (6.6,1);
	\draw (0.5,0.25) .. controls (0.7,0.55) and (1.3,0.55) .. (1.5,0.25);
	\draw (2.5,0.25) .. controls (2.7,0.55) and (3.3,0.55) .. (3.5,0.25);
	\draw (5.5,0.25) .. controls (5.7,0.55) and (6.3,0.55) .. (6.5,0.25);
	\draw[thick,->] (1,0.25) -- (1.3,0.75);
	\draw[thick,->] (3,0.25) -- (3.3,0.75);
	\draw[thick,->] (6.0,0.25) -- (6.3,0.75);
	
	\end{tikzpicture}
\end{center}
\caption{Schematic representation of scheme for the detection of SEE. The system is prepared in each of
its pointer states consecutively and allowed to evolve for time $\tau$. Afterwards the same 
superposition state is excited and the time-dependence of system-coherences is measured.
Any difference in the evolution of coherence for preparation in different pointer states
signifies that a superposition state would be entangled with its environment at time $\tau$
of undisturbed evolution.\label{schem}}
\end{figure}

Otherwise the procedure has to be repeated for a different choice of system pointer state
$|k\rangle$ and again compared with the evolution for $|l\rangle$. Only when all possible values of
$k\neq l$ are exhausted
can one be sure that no entanglement can be witnessed by the procedure.
The procedure is schematically represented
in Fig.~\ref{schem}.

\section{Limitations of applicability \label{sec5}}

The method described above is an entanglement witness  \cite{terhal00,lewenstein00,guhne02,barbieri03}, so a negative result does not 
signify separability. There are two situations when entanglement is present, but cannot
be witnessed here. The first is the same as in the case when the system is a qubit \cite{roszak18},
namely the witness will not detect entanglement if all of the conditional evolution operators
of the environment, $\hat{w}_k(t)$, commute. In this case
the preparation of the environment for time $\tau$ does not change the 
resulting evolution of the system coherences,
which are now always given by
\begin{equation}
\label{jeden}
\rho_{ij}^{(k)}(\tau,t)=c_ic_j^*\mathrm{Tr}\left(
\hat{w}_i(t)\hat{R}(0)
\hat{w}_j^{\dagger}(t)\right).
\end{equation}
This type of entanglement could still be detected by measurements on the environment
since we still have $$\hat{R}_{kk}(t)\neq \hat{R}_{ll}(t)$$ if and only if the state (\ref{sigma}) is entangled, but there is no effect on the evolution of the qudit.

Note that if only some of the conditional evolution operators of the environment mutually commute
then entanglement for some initial states of the system can still be detected, and in many cases
even for all system states. This is because the number of independent criteria (\ref{war1})
is $N-1$ \cite{roszak18} as compared to the $N(N-1)/2$ nontrivial combinations of indices $k$ and $l$. 
If criterion (\ref{war1}) is broken for a given $k$ and $l$ this means that a superposition
with $c_k\neq 0$ and $c_l\neq 0$ will be entangled with its environment at time $\tau$.
This works analogously with indices $k$ and $l'$, but if the criterion is shown to be broken
(using the scheme described in the previous section)
for both sets of indices,
the consequence is that a superposition
with $c_l\neq 0$ and $c_{l'}\neq 0$ will also be entangled with its environment at time $\tau$.
Hence even if $\hat{w}_l(t)$ and $\hat{w}_{l'}(t)$ commute, it is possible to check entanglement
for an initial state with $c_l\neq 0$ and $c_{l'}\neq 0$
using the proposed scheme.

The other situation is when no entanglement of the type witnessed by criterion (\ref{war1})
is generated during the evolution. If only separability criteria of the second type (\ref{war2})
are violated, this type of entanglement does not manifest itself in the conditional evolution
of the environment and cannot be detected using this simple scheme. In fact, detecting 
such entanglement would most likely require tomography of the system-environment state.

\section{Example: NV center spin qutrit \label{sec6}}

\begin{figure}[t]
	\includegraphics[width=0.9\columnwidth]{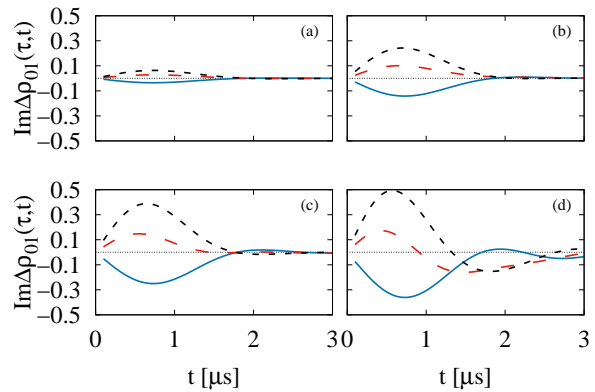}
	\caption{Imaginary part of the evolution post-preparation stage of the difference between a single NV center qutrit coherence
		for different pointer state in the preparation stage, 
		$\rho_{01}^{(k)}(\tau,t)-\rho_{01}^{(q)}(\tau,t)$. Red dashed lines: $k=0$, $q=1$;
		blue solid lines:  $k=0$, $q=-1$; green dashed lines:  $k=-1$, $q=1$.
		The preparation stage lasted for $\tau = 3 \mu$s.
		Applied magnetic field $B_z=0.02$ T. Details of the coupling are contained
		in table \ref{table1}. Different panels correspond to different initial polarizations 
		of the environment: (a) $p=0.1$, (b) $p=0.4$, (c) $p=0.7$, (d) $p=1$.  
	}\label{fig1}
\end{figure}

\begin{figure}[t]
\includegraphics[width=0.9\columnwidth]{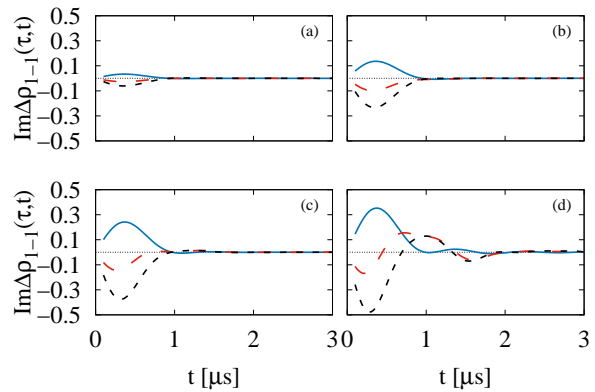}
\caption{As Fig.~\ref{fig1}, but for a different qutrit coherence, 
	$\rho_{1-1}^{(k)}(\tau,t)-\rho_{1-1}^{(q)}(\tau,t)$.
}\label{fig3}
\end{figure}

To exemplify the operation of the scheme described above, we will use it to detect 
entanglement between an NV center spin interacting with an environment of partially
polarized nuclear spins of the spinful carbon
isotope $^{13}$C in the diamond lattice \cite{doherty13,wood18,awschalom18,tchebotareva19}. The dominant carbon isotope $^{12}$C is spinless
and does not contribute to NV center spin decoherence, so that the environment is sparse.
The lowest energy level of the NV center is effectively a spin qutrit, with $S=1$, so
the dimension of the system is $N=3$
and only two entanglement criteria of the first type (\ref{war1})
need to be checked to show that entanglement would be present for any initial superposition
state of the system. 

For the majority of values of the applied magnetic field, the pure dephasing approximation
can be used to describe this system and environment \cite{zhao12,kwiatkowski18}, so the Hamiltonian
is of the form given by eq.~(\ref{ham0}). For convenience we will change the summation over system
states to $k = -1,0,1$, so the index $k$ corresponds to the three lowest level spin states 
of the qutrit which are also its pointer states \cite{zurek81,zurek03}. The energies 
which describe the free evolution of the qutrit in the Hamiltonian (\ref{ham0}) are equal to
$\varepsilon_0=0$ and
$\varepsilon_{\pm 1}=\Delta \pm\gamma_{e}B_{z}$.
	Here $B_{z}$ is the magnetic field applied in the $z$ direction, so $\gamma_{e}B_{z}$
	is the magnetic-field induced splitting of the qutrit levels
	($\gamma_e=28.08$ MHz/T is the electron gyromagnetic ratio). The zero-field splitting,
	$\Delta\hat{S}_z^2$ with $\Delta = 2.87$ GHz, determines the $z$ direction, which is dependent on the geometry of the NV center.
	The term is responsible for the uneven energy splitting of the qutrit
	states.
	The free evolution of the environment is given by
	$
	\hat{H}_{\mathrm{E}}=\sum_{j}\gamma_{n}B_{z}\hat{I}^{z}_{j}$,
where $j$ labels the $^{13}$C spins, $\gamma_{n} \! = \! 10.71$ MHz/T
is the gyromagnetic ratio for $^{13}$C nuclei, and $\hat{I}^{z}_{j}$ are operators of the $z$ component of the nuclear spins.

\begin{table}
	\begin{tabular}{|c|c|c|c|c|}\hline
		$k$&$r_k$ [nm]&$\mathbb{A}^{z,x}_k$ [1/$\mu$s]&$\mathbb{A}^{z,y}_k$ [1/$\mu$s]&$\mathbb{A}^{z,z}_k$ [1/$\mu$s]\\
		\hline\hline
		$1$&$0.504422$&	$1.37617$&	$0$&	$0.973096$\\
		\hline
		$2$&$0.563961$&	$0.196941$&	$	0.682223$&	$	-0.417774$\\
		\hline
		$3$&$0.563961$&	$	-0.689293$&	$	0.170556$&	$	-0.417774$\\
		\hline
		$4$&$0.563961$&	$	0.492352$&	$	-0.511667$&	$	-0.417774$\\
		\hline
		$5$&$0.617788$&	$	0.499393$&	$	0$&	$	-0.353124$\\
		\hline
		$6$&$0.636801$&	$	0.469395$&	$	-0.487809$&	$	-0.0189664$\\
		\hline
		$7$&$0.636801$&	$	-0.0134113$&	$	-0.116145$&	$	-0.47416$\\
		\hline
		$8$&$0.667287$&	$	-0.297224$&	 $	0.220631$&	$	-0.300241$\\
		\hline
		$9$&$0.667287$&	$	-0.169842$&	 $	0.58835	$&	$	0.0600483$\\
		\hline
		$10$&$0.667287$&	$	-0.382145$&	 $	0.220631$&	$	0.660531$\\
		\hline
		$11$&$0.667287$&	$	0$&	$	0$&	 $-0.420338$\\
		\hline
		$12$&$0.684928$&	$	0.326087$&	$	0.242057$&	$	-0.22399$\\
		\hline
		$13$&$0.684928$&	$	0.251553$&	$	-0.0484114$&	$	-0.329398$\\
		\hline
		$14$&$0.684928$&	$	-0.372671$&	 $	0.161371$&	$	-0.22399$\\
		\hline
	\end{tabular}
	\caption{Table of calculated coupling constants for fourteen environmental spins at randomly generated locations
		around the NV center.  \label{table1}}
\end{table}
 
 The hyperfine interaction between the qutrit and the nuclear environment yields 
 operators which describe the response of the environment to a given pointer state of the 
 qutrit. They are given by $\hat{V}_0=0$ and $\hat{V}_{\pm 1}=\pm\hat{V}$, with
\begin{equation}
	\label{v}
	\hat{V} = \sum_{j}\left( \mathbb{A}^{z,x}_j\hat{I}^{x}_{j}
	+\mathbb{A}^{z,y}_j\hat{I}^{y}_{j}+\mathbb{A}^{z,z}_j\hat{I}^{z}_{j}
	\right).
	\end{equation}
Here the coupling constants for each direction are of the form
\begin{equation}
	\label{a}
	\mathbb{A}^{z,i}_j=\frac{\mu_0}{4\pi}\frac{\gamma_e\gamma_n}{r_{j}^3}
	\left(1-\frac{3(\mathbf{r}_{j}\cdot\hat{\mathbf{i}})(\mathbf{r}_{j}\cdot\hat{\mathbf{z}})}{r_{j}^2}
	\right),
	\end{equation}
where $\mathbf{r}_{j}$ is a displacement vector between the $j$-th nucleus and the qutrit, while
	$\hat{\mathbf{i}}=\hat{\mathbf{x}},\hat{\mathbf{y}},\hat{\mathbf{z}}$
	are unit vectors in three distinct directions. $\mu_0$ is the magnetic permeability of the vacuum.
	
The conditional evolution operators of the environment (\ref{w0}) which enter 
the full system-environment evolution operator (\ref{u}) are now straightforward to compute
(see Ref.~\cite{strzalka20} for details) and are given by
		\begin{eqnarray}
		\hat{w}_{\pm 1}(t)&=&\bigotimes_j\left[\cos(M^j_{\pm}t)\unit_j \right.\\
\nonumber
&&-\left.\frac{i  \sin(M^j_{\pm}t)}{M^j_{\pm}}\left(\pm \mathbb{A}^{z,\mathrm{pl}}_j\hat{I}^{x}_{j}+
\left(\gamma_{e}B_{z}\pm \mathbb{A}^{z,z}_j\right)\hat{I}^{z}_{j}\right)\right],\\
		\hat{w}_0(t)&=&\bigotimes_j\left[\cos(\gamma_{e}B_{z}t)\unit_j-i \sin(\gamma_{e}B_{z}t)\hat{I}^{z}_{j}\right],
		\end{eqnarray}
with $
M_{\pm}= \sqrt{A_x^2+(\gamma_{n}B_{z}\pm A_z)^2}$ and
$\mathbb{A}^{z,\mathrm{pl}}_j=\sqrt{(\mathbb{A}^{z,y}_j)^2+(\mathbb{A}^{z,y}_j)^2}$.

Given the initial state of the environment, we can find the evolution of the coherences
which are needed to detect QEE,
namely eq.~(\ref{rhoij}).
The thermal-equilibrium state of this environment (with respect to its free Hamiltonian) is 
effectively proportional to unity due to the small value of the gyromagnetic ratio for $^{13}$C nuclei.
As such a state will not lead to entanglement, we will consider a dynamically polarized nuclear
environment \cite{fischer13,london13,hovav18}, so that
\begin{equation}
	\hat{R}(0)  =  \bigotimes_{j} \frac{1}{2}(\unit_j + p_{j}\hat{I}^{z}_{j}),
\label{env_ini}
\end{equation} 
where $p_j \! \in \! [-1,1]$ is the polarization of the $j$-th nucleus.

In Figs \ref{fig1} and \ref{fig3} we plot the difference of the evolution
of a single chosen 
coherence of the qutrit (as a function of time $t$) for different pointer states in the preparation part of the procedure
(up to time $\tau$), $\rho_{ij}^{(k)}(\tau,t)-\rho_{ij}^{(q)}(\tau,t)$.
Each figure contains curves corresponding to all three combinations of pointer states
in the preparation stage, $k,q=-1,0,1$ (any two would suffice to determine if entanglement 
would be present for any initial system superposition state
after time $\tau$). Fig.~\ref{fig1} shows the difference in evolution 
for the coherence between the $|0\rangle$ and $|1\rangle$ qutrit states,
while Fig.~\ref{fig3} for the coherence between the $|-1\rangle$ and $|1\rangle$ states.
The evolution of the third coherence is not shown as it would be superfluous. 
Furthermore, we only show the imaginary part of the difference of the evolution,
because the results are more striking in this case, and the real part would not 
bring anything relevant to the discussion (since the scheme has already witnessed entanglement).

The preparation time $\tau=3\mu$s was chosen long so that the presence of the qutrit in a pointer
state has the strongest possible effect on the new (post-preparation) state of the environment
and consequently the states differ most notably for different pointer states.
This in turn enhances the differences observed in qutrit evolution.
The magnetic field is $B_z=0.02$ T. 
Each plot contains four panels corresponding to four different initial states of the environment,
characterized by different polarizations. As in the NV-center spin qubit case \cite{roszak19a}, an unpolarized
environment does not entangle with the qutrit and the magnitude of the observed effect grows
with higher initial polarization.
The results are given for an environment consisting of fourteen nuclear spins placed
at randomly generated locations. Table \ref{table1} contains the distances between each nuclear spin
and the NV center, as well as of the coupling constants used, which were calculated using
eq.~(\ref{a}).

\section{Conclusion \label{sec7}}

We have proposed a scheme for the indirect detection of entanglement between a system of any dimensionality with an environment interacting via a Hamilotnian which leads to pure dephasing of
the qudit. This is a generalization of a scheme proposed for a system composed of a single qubit
\cite{roszak19a},
but even though the number of separability criteria grows quadratically with the size of the system
\cite{roszak18},
the complexity of the scheme only grows linearly. The price to pay is that the set of 
states for which entanglement cannot be detected using the scheme is also larger, and entanglement
connected with braking separability criteria based on commutation between products of 
conditional evolution operators of the environment (which does not exist for a qubit system)
cannot be detected.

On the other hand, the scheme only requires 
straightforward operations and measurements on the system of interest and allows for detection 
of entanglement in systems too large for any type of state tomography to be feasible. It detects
entanglement which manifests itself in the evolution of the environment, and as such is most
likely to have an effect on the evolution of the system. The mechanism of the scheme directly
relies on the influence of SEE on the evolution of the system, so
it will detect the type of entanglement which is bound to be most detrimental to the system
(or description of system evolution which assumes separability, such as using quantum channels \cite{nielsen00}). 

The authors would like to thank Łukasz Cywiński and Damian Kwiatkowski for sharing the data contained
in Table \ref{table1}.


\begin{thebibliography}{43}%
	\makeatletter
	\providecommand \@ifxundefined [1]{%
		\@ifx{#1\undefined}
	}%
	\providecommand \@ifnum [1]{%
		\ifnum #1\expandafter \@firstoftwo
		\else \expandafter \@secondoftwo
		\fi
	}%
	\providecommand \@ifx [1]{%
		\ifx #1\expandafter \@firstoftwo
		\else \expandafter \@secondoftwo
		\fi
	}%
	\providecommand \natexlab [1]{#1}%
	\providecommand \enquote  [1]{``#1''}%
	\providecommand \bibnamefont  [1]{#1}%
	\providecommand \bibfnamefont [1]{#1}%
	\providecommand \citenamefont [1]{#1}%
	\providecommand \href@noop [0]{\@secondoftwo}%
	\providecommand \href [0]{\begingroup \@sanitize@url \@href}%
	\providecommand \@href[1]{\@@startlink{#1}\@@href}%
	\providecommand \@@href[1]{\endgroup#1\@@endlink}%
	\providecommand \@sanitize@url [0]{\catcode `\\12\catcode `\$12\catcode
		`\&12\catcode `\#12\catcode `\^12\catcode `\_12\catcode `\%12\relax}%
	\providecommand \@@startlink[1]{}%
	\providecommand \@@endlink[0]{}%
	\providecommand \url  [0]{\begingroup\@sanitize@url \@url }%
	\providecommand \@url [1]{\endgroup\@href {#1}{\urlprefix }}%
	\providecommand \urlprefix  [0]{URL }%
	\providecommand \Eprint [0]{\href }%
	\providecommand \doibase [0]{https://doi.org/}%
	\providecommand \selectlanguage [0]{\@gobble}%
	\providecommand \bibinfo  [0]{\@secondoftwo}%
	\providecommand \bibfield  [0]{\@secondoftwo}%
	\providecommand \translation [1]{[#1]}%
	\providecommand \BibitemOpen [0]{}%
	\providecommand \bibitemStop [0]{}%
	\providecommand \bibitemNoStop [0]{.\EOS\space}%
	\providecommand \EOS [0]{\spacefactor3000\relax}%
	\providecommand \BibitemShut  [1]{\csname bibitem#1\endcsname}%
	\let\auto@bib@innerbib\@empty
	\bibitem [{\citenamefont {Nielsen}\ and\ \citenamefont
		{Chuang}(2000)}]{nielsen00}%
	\BibitemOpen
	\bibfield  {author} {\bibinfo {author} {\bibfnamefont {M.~A.}\ \bibnamefont
			{Nielsen}}\ and\ \bibinfo {author} {\bibfnamefont {I.~L.}\ \bibnamefont
			{Chuang}},\ }\href@noop {} {\emph {\bibinfo {title} {Quantum Computation and
				Quantum Information}}}\ (\bibinfo  {publisher} {Cambridge University Press},\
	\bibinfo {address} {Cambridge},\ \bibinfo {year} {2000})\BibitemShut
	{NoStop}%
	\bibitem [{\citenamefont {Plenio}\ and\ \citenamefont
		{Virmani}(2007)}]{plenio07}%
	\BibitemOpen
	\bibfield  {author} {\bibinfo {author} {\bibfnamefont {M.~B.}\ \bibnamefont
			{Plenio}}\ and\ \bibinfo {author} {\bibfnamefont {S.}~\bibnamefont
			{Virmani}},\ }\bibfield  {title} {\bibinfo {title} {An introduction to
			entanglement measures},\ }\href
	{http://www.rintonpress.com/xqic7/qic-7-12/001-051.pdf} {\bibfield  {journal}
		{\bibinfo  {journal} {Quantum Inf. Comput.}\ }\textbf {\bibinfo {volume}
			{7}},\ \bibinfo {pages} {1} (\bibinfo {year} {2007})}\BibitemShut {NoStop}%
	\bibitem [{\citenamefont {Wootters}(1998)}]{wootters98}%
	\BibitemOpen
	\bibfield  {author} {\bibinfo {author} {\bibfnamefont {W.~K.}\ \bibnamefont
			{Wootters}},\ }\bibfield  {title} {\bibinfo {title} {Entanglement of
			formation of an arbitrary state of two qubits},\ }\href@noop {} {\ \textbf
		{\bibinfo {volume} {80}},\ \bibinfo {pages} {2245} (\bibinfo {year}
		{1998})}\BibitemShut {NoStop}%
	\bibitem [{\citenamefont {Horodecki}\ \emph {et~al.}(2009)\citenamefont
		{Horodecki}, \citenamefont {Horodecki}, \citenamefont {Horodecki},\ and\
		\citenamefont {Horodecki}}]{horodecki09}%
	\BibitemOpen
	\bibfield  {author} {\bibinfo {author} {\bibfnamefont {R.}~\bibnamefont
			{Horodecki}}, \bibinfo {author} {\bibfnamefont {P.}~\bibnamefont
			{Horodecki}}, \bibinfo {author} {\bibfnamefont {M.}~\bibnamefont
			{Horodecki}},\ and\ \bibinfo {author} {\bibfnamefont {K.}~\bibnamefont
			{Horodecki}},\ }\bibfield  {title} {\bibinfo {title} {Quantum entanglement},\
	}\href {https://doi.org/10.1103/RevModPhys.81.865} {\bibfield  {journal}
		{\bibinfo  {journal} {Rev. Mod. Phys.}\ }\textbf {\bibinfo {volume} {81}},\
		\bibinfo {pages} {865} (\bibinfo {year} {2009})}\BibitemShut {NoStop}%
	\bibitem [{\citenamefont {Vidal}\ and\ \citenamefont {Werner}(2002)}]{vidal02}%
	\BibitemOpen
	\bibfield  {author} {\bibinfo {author} {\bibfnamefont {G.}~\bibnamefont
			{Vidal}}\ and\ \bibinfo {author} {\bibfnamefont {R.~F.}\ \bibnamefont
			{Werner}},\ }\bibfield  {title} {\bibinfo {title} {Computable measure of
			entanglement},\ }\href {https://doi.org/10.1103/PhysRevA.65.032314}
	{\bibfield  {journal} {\bibinfo  {journal} {Phys. Rev. A}\ }\textbf {\bibinfo
			{volume} {65}},\ \bibinfo {pages} {032314} (\bibinfo {year}
		{2002})}\BibitemShut {NoStop}%
	\bibitem [{\citenamefont {Lee}\ \emph {et~al.}(2000)\citenamefont {Lee},
		\citenamefont {Kim}, \citenamefont {Park},\ and\ \citenamefont
		{Lee}}]{lee00a}%
	\BibitemOpen
	\bibfield  {author} {\bibinfo {author} {\bibfnamefont {J.}~\bibnamefont
			{Lee}}, \bibinfo {author} {\bibfnamefont {M.}~\bibnamefont {Kim}}, \bibinfo
		{author} {\bibfnamefont {Y.}~\bibnamefont {Park}},\ and\ \bibinfo {author}
		{\bibfnamefont {S.}~\bibnamefont {Lee}},\ }\bibfield  {title} {\bibinfo
		{title} {Partial teleportation of entanglement in a noisy environment},\
	}\href@noop {} {\bibfield  {journal} {\bibinfo  {journal} {J. Mod. Opt.}\
		}\textbf {\bibinfo {volume} {47}},\ \bibinfo {pages} {2157} (\bibinfo {year}
		{2000})}\BibitemShut {NoStop}%
	\bibitem [{\citenamefont {Plenio}(2005)}]{plenio05b}%
	\BibitemOpen
	\bibfield  {author} {\bibinfo {author} {\bibfnamefont {M.~B.}\ \bibnamefont
			{Plenio}},\ }\bibfield  {title} {\bibinfo {title} {Logarithmic negativity: A
			full entanglement monotone that is not convex},\ }\href
	{https://doi.org/10.1103/PhysRevLett.95.090503} {\bibfield  {journal}
		{\bibinfo  {journal} {Phys. Rev. Lett.}\ }\textbf {\bibinfo {volume} {95}},\
		\bibinfo {pages} {090503} (\bibinfo {year} {2005})}\BibitemShut {NoStop}%
	\bibitem [{\citenamefont {Nakano}\ \emph {et~al.}(2013)\citenamefont {Nakano},
		\citenamefont {Piani},\ and\ \citenamefont {Adesso}}]{nakano13}%
	\BibitemOpen
	\bibfield  {author} {\bibinfo {author} {\bibfnamefont {T.}~\bibnamefont
			{Nakano}}, \bibinfo {author} {\bibfnamefont {M.}~\bibnamefont {Piani}},\ and\
		\bibinfo {author} {\bibfnamefont {G.}~\bibnamefont {Adesso}},\ }\bibfield
	{title} {\bibinfo {title} {Negativity of quantumness and its
			interpretations},\ }\href {https://doi.org/10.1103/PhysRevA.88.012117}
	{\bibfield  {journal} {\bibinfo  {journal} {Phys. Rev. A}\ }\textbf {\bibinfo
			{volume} {88}},\ \bibinfo {pages} {012117} (\bibinfo {year}
		{2013})}\BibitemShut {NoStop}%
	\bibitem [{\citenamefont {Horodecki}(1997)}]{horodecki97b}%
	\BibitemOpen
	\bibfield  {author} {\bibinfo {author} {\bibfnamefont {P.}~\bibnamefont
			{Horodecki}},\ }\bibfield  {title} {\bibinfo {title} {Separability criterion
			and inseparable mixed states with positive partial transposition},\ }\href
	{https://doi.org/https://doi.org/10.1016/S0375-9601(97)00416-7} {\bibfield
		{journal} {\bibinfo  {journal} {Physics Letters A}\ }\textbf {\bibinfo
			{volume} {232}},\ \bibinfo {pages} {333} (\bibinfo {year}
		{1997})}\BibitemShut {NoStop}%
	\bibitem [{\citenamefont {Horodecki}\ \emph {et~al.}(1998)\citenamefont
		{Horodecki}, \citenamefont {Horodecki},\ and\ \citenamefont
		{Horodecki}}]{horodecki98}%
	\BibitemOpen
	\bibfield  {author} {\bibinfo {author} {\bibfnamefont {M.}~\bibnamefont
			{Horodecki}}, \bibinfo {author} {\bibfnamefont {P.}~\bibnamefont
			{Horodecki}},\ and\ \bibinfo {author} {\bibfnamefont {R.}~\bibnamefont
			{Horodecki}},\ }\bibfield  {title} {\bibinfo {title} {Mixed-state
			entanglement and distillation: Is there a ``bound'' entanglement in
			nature?},\ }\href {https://doi.org/10.1103/PhysRevLett.80.5239} {\bibfield
		{journal} {\bibinfo  {journal} {Phys. Rev. Lett.}\ }\textbf {\bibinfo
			{volume} {80}},\ \bibinfo {pages} {5239} (\bibinfo {year}
		{1998})}\BibitemShut {NoStop}%
	\bibitem [{\citenamefont {Peres}(1996)}]{peres96a}%
	\BibitemOpen
	\bibfield  {author} {\bibinfo {author} {\bibfnamefont {A.}~\bibnamefont
			{Peres}},\ }\bibfield  {title} {\bibinfo {title} {Separability criterion for
			density matrices},\ }\href {https://doi.org/10.1103/PhysRevLett.77.1413}
	{\bibfield  {journal} {\bibinfo  {journal} {Phys. Rev. Lett.}\ }\textbf
		{\bibinfo {volume} {77}},\ \bibinfo {pages} {1413} (\bibinfo {year}
		{1996})}\BibitemShut {NoStop}%
	\bibitem [{\citenamefont {Horodecki}\ \emph {et~al.}(1996)\citenamefont
		{Horodecki}, \citenamefont {Horodecki},\ and\ \citenamefont
		{Horodecki}}]{horodecki96}%
	\BibitemOpen
	\bibfield  {author} {\bibinfo {author} {\bibfnamefont {M.}~\bibnamefont
			{Horodecki}}, \bibinfo {author} {\bibfnamefont {P.}~\bibnamefont
			{Horodecki}},\ and\ \bibinfo {author} {\bibfnamefont {R.}~\bibnamefont
			{Horodecki}},\ }\bibfield  {title} {\bibinfo {title} {Separability of mixed
			states: necessary and sufficient conditions},\ }\href
	{https://doi.org/https://doi.org/10.1016/S0375-9601(96)00706-2} {\bibfield
		{journal} {\bibinfo  {journal} {Physics Letters A}\ }\textbf {\bibinfo
			{volume} {223}},\ \bibinfo {pages} {1} (\bibinfo {year} {1996})}\BibitemShut
	{NoStop}%
	\bibitem [{\citenamefont {Roszak}\ and\ \citenamefont
		{Cywi\ifmmode~\acute{n}\else \'{n}\fi{}ski}(2015)}]{roszak15}%
	\BibitemOpen
	\bibfield  {author} {\bibinfo {author} {\bibfnamefont {K.}~\bibnamefont
			{Roszak}}\ and\ \bibinfo {author} {\bibfnamefont {L.}~\bibnamefont
			{Cywi\ifmmode~\acute{n}\else \'{n}\fi{}ski}},\ }\bibfield  {title} {\bibinfo
		{title} {Characterization and measurement of qubit-environment-entanglement
			generation during pure dephasing},\ }\href
	{https://doi.org/10.1103/PhysRevA.92.032310} {\bibfield  {journal} {\bibinfo
			{journal} {Phys. Rev. A}\ }\textbf {\bibinfo {volume} {92}},\ \bibinfo
		{pages} {032310} (\bibinfo {year} {2015})}\BibitemShut {NoStop}%
	\bibitem [{\citenamefont {Roszak}(2018)}]{roszak18}%
	\BibitemOpen
	\bibfield  {author} {\bibinfo {author} {\bibfnamefont {K.}~\bibnamefont
			{Roszak}},\ }\bibfield  {title} {\bibinfo {title} {Criteria for
			system-environment entanglement generation for systems of any size in
			pure-dephasing evolutions},\ }\href
	{https://doi.org/10.1103/PhysRevA.98.052344} {\bibfield  {journal} {\bibinfo
			{journal} {Phys. Rev. A}\ }\textbf {\bibinfo {volume} {98}},\ \bibinfo
		{pages} {052344} (\bibinfo {year} {2018})}\BibitemShut {NoStop}%
	\bibitem [{\citenamefont {Roszak}(2020)}]{roszak20}%
	\BibitemOpen
	\bibfield  {author} {\bibinfo {author} {\bibfnamefont {K.}~\bibnamefont
			{Roszak}},\ }\bibfield  {title} {\bibinfo {title} {Measure of
			qubit-environment entanglement for pure dephasing evolutions},\ }\href
	{https://doi.org/10.1103/PhysRevResearch.2.043062} {\bibfield  {journal}
		{\bibinfo  {journal} {Phys. Rev. Research}\ }\textbf {\bibinfo {volume}
			{2}},\ \bibinfo {pages} {043062} (\bibinfo {year} {2020})}\BibitemShut
	{NoStop}%
	\bibitem [{\citenamefont {Strza\l{}ka}\ \emph {et~al.}(2020)\citenamefont
		{Strza\l{}ka}, \citenamefont {Kwiatkowski}, \citenamefont
		{Cywi\ifmmode~\acute{n}\else \'{n}\fi{}ski},\ and\ \citenamefont
		{Roszak}}]{strzalka20}%
	\BibitemOpen
	\bibfield  {author} {\bibinfo {author} {\bibfnamefont {M.}~\bibnamefont
			{Strza\l{}ka}}, \bibinfo {author} {\bibfnamefont {D.}~\bibnamefont
			{Kwiatkowski}}, \bibinfo {author} {\bibfnamefont {L.}~\bibnamefont
			{Cywi\ifmmode~\acute{n}\else \'{n}\fi{}ski}},\ and\ \bibinfo {author}
		{\bibfnamefont {K.}~\bibnamefont {Roszak}},\ }\bibfield  {title} {\bibinfo
		{title} {Qubit-environment negativity versus fidelity of conditional
			environmental states for a nitrogen-vacancy-center spin qubit interacting
			with a nuclear environment},\ }\href
	{https://doi.org/10.1103/PhysRevA.102.042602} {\bibfield  {journal} {\bibinfo
			{journal} {Phys. Rev. A}\ }\textbf {\bibinfo {volume} {102}},\ \bibinfo
		{pages} {042602} (\bibinfo {year} {2020})}\BibitemShut {NoStop}%
	\bibitem [{\citenamefont {Roszak}\ and\ \citenamefont
		{Cywi\ifmmode~\acute{n}\else \'{n}\fi{}ski}(2018)}]{roszak17}%
	\BibitemOpen
	\bibfield  {author} {\bibinfo {author} {\bibfnamefont {K.}~\bibnamefont
			{Roszak}}\ and\ \bibinfo {author} {\bibfnamefont {L.}~\bibnamefont
			{Cywi\ifmmode~\acute{n}\else \'{n}\fi{}ski}},\ }\bibfield  {title} {\bibinfo
		{title} {Equivalence of qubit-environment entanglement and discord generation
			via pure dephasing interactions and the resulting consequences},\ }\href
	{https://doi.org/10.1103/PhysRevA.97.012306} {\bibfield  {journal} {\bibinfo
			{journal} {Phys. Rev. A}\ }\textbf {\bibinfo {volume} {97}},\ \bibinfo
		{pages} {012306} (\bibinfo {year} {2018})}\BibitemShut {NoStop}%
	\bibitem [{\citenamefont {Roszak}\ \emph {et~al.}(2015)\citenamefont {Roszak},
		\citenamefont {Filip},\ and\ \citenamefont {Novotný}}]{roszak15b}%
	\BibitemOpen
	\bibfield  {author} {\bibinfo {author} {\bibfnamefont {K.}~\bibnamefont
			{Roszak}}, \bibinfo {author} {\bibfnamefont {R.}~\bibnamefont {Filip}},\ and\
		\bibinfo {author} {\bibfnamefont {T.}~\bibnamefont {Novotný}},\ }\bibfield
	{title} {\bibinfo {title} {Decoherence control by quantum decoherence
			itself.},\ }\href {https://doi.org/10.1038/srep09796} {\bibfield  {journal}
		{\bibinfo  {journal} {Scientific Reports}\ }\textbf {\bibinfo {volume} {5}},\
		\bibinfo {pages} {9796} (\bibinfo {year} {2015})}\BibitemShut {NoStop}%
	\bibitem [{\citenamefont {Roszak}\ \emph {et~al.}(2019)\citenamefont {Roszak},
		\citenamefont {Kwiatkowski},\ and\ \citenamefont {Cywi\ifmmode~\acute{n}\else
			\'{n}\fi{}ski}}]{roszak19a}%
	\BibitemOpen
	\bibfield  {author} {\bibinfo {author} {\bibfnamefont {K.}~\bibnamefont
			{Roszak}}, \bibinfo {author} {\bibfnamefont {D.}~\bibnamefont
			{Kwiatkowski}},\ and\ \bibinfo {author} {\bibfnamefont {L.}~\bibnamefont
			{Cywi\ifmmode~\acute{n}\else \'{n}\fi{}ski}},\ }\bibfield  {title} {\bibinfo
		{title} {How to detect qubit-environment entanglement generated during qubit
			dephasing},\ }\href {https://doi.org/10.1103/PhysRevA.100.022318} {\bibfield
		{journal} {\bibinfo  {journal} {Phys. Rev. A}\ }\textbf {\bibinfo {volume}
			{100}},\ \bibinfo {pages} {022318} (\bibinfo {year} {2019})}\BibitemShut
	{NoStop}%
	\bibitem [{\citenamefont {Rzepkowski}\ and\ \citenamefont
		{Roszak}(2020)}]{rzepkowski20}%
	\BibitemOpen
	\bibfield  {author} {\bibinfo {author} {\bibfnamefont {B.}~\bibnamefont
			{Rzepkowski}}\ and\ \bibinfo {author} {\bibfnamefont {K.}~\bibnamefont
			{Roszak}},\ }\href@noop {} {\bibinfo {title} {A scheme for direct detection
			of qubit-environment entanglement generated during qubit pure dephasing}}
	(\bibinfo {year} {2020}),\ \Eprint {https://arxiv.org/abs/2002.10901}
	{arXiv:2002.10901 [quant-ph]} \BibitemShut {NoStop}%
	\bibitem [{\citenamefont {Medford}\ \emph {et~al.}(2012)\citenamefont
		{Medford}, \citenamefont {Cywi\ifmmode~\acute{n}\else \'{n}\fi{}ski},
		\citenamefont {Barthel}, \citenamefont {Marcus}, \citenamefont {Hanson},\
		and\ \citenamefont {Gossard}}]{medford12}%
	\BibitemOpen
	\bibfield  {author} {\bibinfo {author} {\bibfnamefont {J.}~\bibnamefont
			{Medford}}, \bibinfo {author} {\bibfnamefont {L.}~\bibnamefont
			{Cywi\ifmmode~\acute{n}\else \'{n}\fi{}ski}}, \bibinfo {author}
		{\bibfnamefont {C.}~\bibnamefont {Barthel}}, \bibinfo {author} {\bibfnamefont
			{C.~M.}\ \bibnamefont {Marcus}}, \bibinfo {author} {\bibfnamefont {M.~P.}\
			\bibnamefont {Hanson}},\ and\ \bibinfo {author} {\bibfnamefont {A.~C.}\
			\bibnamefont {Gossard}},\ }\bibfield  {title} {\bibinfo {title} {Scaling of
			dynamical decoupling for spin qubits},\ }\href
	{https://doi.org/10.1103/PhysRevLett.108.086802} {\bibfield  {journal}
		{\bibinfo  {journal} {Phys. Rev. Lett.}\ }\textbf {\bibinfo {volume} {108}},\
		\bibinfo {pages} {086802} (\bibinfo {year} {2012})}\BibitemShut {NoStop}%
	\bibitem [{\citenamefont {Kawakami}\ \emph {et~al.}(2016)\citenamefont
		{Kawakami}, \citenamefont {Jullien}, \citenamefont {Scarlino}, \citenamefont
		{Ward}, \citenamefont {Savage}, \citenamefont {Lagally}, \citenamefont
		{Dobrovitski}, \citenamefont {Friesen}, \citenamefont {Coppersmith},
		\citenamefont {Eriksson},\ and\ \citenamefont {Vandersypen}}]{kawakami16}%
	\BibitemOpen
	\bibfield  {author} {\bibinfo {author} {\bibfnamefont {E.}~\bibnamefont
			{Kawakami}}, \bibinfo {author} {\bibfnamefont {T.}~\bibnamefont {Jullien}},
		\bibinfo {author} {\bibfnamefont {P.}~\bibnamefont {Scarlino}}, \bibinfo
		{author} {\bibfnamefont {D.~R.}\ \bibnamefont {Ward}}, \bibinfo {author}
		{\bibfnamefont {D.~E.}\ \bibnamefont {Savage}}, \bibinfo {author}
		{\bibfnamefont {M.~G.}\ \bibnamefont {Lagally}}, \bibinfo {author}
		{\bibfnamefont {V.~V.}\ \bibnamefont {Dobrovitski}}, \bibinfo {author}
		{\bibfnamefont {M.}~\bibnamefont {Friesen}}, \bibinfo {author} {\bibfnamefont
			{S.~N.}\ \bibnamefont {Coppersmith}}, \bibinfo {author} {\bibfnamefont
			{M.~A.}\ \bibnamefont {Eriksson}},\ and\ \bibinfo {author} {\bibfnamefont
			{L.~M.~K.}\ \bibnamefont {Vandersypen}},\ }\bibfield  {title} {\bibinfo
		{title} {Gate fidelity and coherence of an electron spin in an si/sige
			quantum dot with micromagnet},\ }\href
	{https://doi.org/10.1073/pnas.1603251113} {\bibfield  {journal} {\bibinfo
			{journal} {Proceedings of the National Academy of Sciences}\ }\textbf
		{\bibinfo {volume} {113}},\ \bibinfo {pages} {11738} (\bibinfo {year}
		{2016})},\ \Eprint
	{https://arxiv.org/abs/https://www.pnas.org/content/113/42/11738.full.pdf}
	{https://www.pnas.org/content/113/42/11738.full.pdf} \BibitemShut {NoStop}%
	\bibitem [{\citenamefont {Zhao}\ \emph {et~al.}(2012)\citenamefont {Zhao},
		\citenamefont {Ho},\ and\ \citenamefont {Liu}}]{zhao12}%
	\BibitemOpen
	\bibfield  {author} {\bibinfo {author} {\bibfnamefont {N.}~\bibnamefont
			{Zhao}}, \bibinfo {author} {\bibfnamefont {S.-W.}\ \bibnamefont {Ho}},\ and\
		\bibinfo {author} {\bibfnamefont {R.-B.}\ \bibnamefont {Liu}},\ }\bibfield
	{title} {\bibinfo {title} {Decoherence and dynamical decoupling control of
			nitrogen vacancy center electron spins in nuclear spin baths},\ }\href
	{https://doi.org/10.1103/PhysRevB.85.115303} {\bibfield  {journal} {\bibinfo
			{journal} {Phys. Rev. B}\ }\textbf {\bibinfo {volume} {85}},\ \bibinfo
		{pages} {115303} (\bibinfo {year} {2012})}\BibitemShut {NoStop}%
	\bibitem [{\citenamefont {Kwiatkowski}\ and\ \citenamefont
		{Cywi\ifmmode~\acute{n}\else \'{n}\fi{}ski}(2018)}]{kwiatkowski18}%
	\BibitemOpen
	\bibfield  {author} {\bibinfo {author} {\bibfnamefont {D.}~\bibnamefont
			{Kwiatkowski}}\ and\ \bibinfo {author} {\bibfnamefont {L.}~\bibnamefont
			{Cywi\ifmmode~\acute{n}\else \'{n}\fi{}ski}},\ }\bibfield  {title} {\bibinfo
		{title} {Decoherence of two entangled spin qubits coupled to an interacting
			sparse nuclear spin bath: Application to nitrogen vacancy centers},\ }\href
	{https://doi.org/10.1103/PhysRevB.98.155202} {\bibfield  {journal} {\bibinfo
			{journal} {Phys. Rev. B}\ }\textbf {\bibinfo {volume} {98}},\ \bibinfo
		{pages} {155202} (\bibinfo {year} {2018})}\BibitemShut {NoStop}%
	\bibitem [{\citenamefont {Malinowski}\ \emph {et~al.}(2017)\citenamefont
		{Malinowski}, \citenamefont {Martins}, \citenamefont {Nissen}, \citenamefont
		{Barnes}, \citenamefont {Cywiński}, \citenamefont {Rudner}, \citenamefont
		{Fallahi}, \citenamefont {Gardner}, \citenamefont {Manfra}, \citenamefont
		{Marcus},\ and\ \citenamefont {Kuemmeth}}]{malinowski17}%
	\BibitemOpen
	\bibfield  {author} {\bibinfo {author} {\bibfnamefont {F.~K.}\ \bibnamefont
			{Malinowski}}, \bibinfo {author} {\bibfnamefont {F.}~\bibnamefont {Martins}},
		\bibinfo {author} {\bibfnamefont {P.~D.}\ \bibnamefont {Nissen}}, \bibinfo
		{author} {\bibfnamefont {E.}~\bibnamefont {Barnes}}, \bibinfo {author}
		{\bibfnamefont {{\L}.}~\bibnamefont {Cywiński}}, \bibinfo {author}
		{\bibfnamefont {M.~S.}\ \bibnamefont {Rudner}}, \bibinfo {author}
		{\bibfnamefont {S.}~\bibnamefont {Fallahi}}, \bibinfo {author} {\bibfnamefont
			{G.~C.}\ \bibnamefont {Gardner}}, \bibinfo {author} {\bibfnamefont {M.~J.}\
			\bibnamefont {Manfra}}, \bibinfo {author} {\bibfnamefont {C.~M.}\
			\bibnamefont {Marcus}},\ and\ \bibinfo {author} {\bibfnamefont
			{F.}~\bibnamefont {Kuemmeth}},\ }\bibfield  {title} {\bibinfo {title}
		{Nuclear spins in gallium arsenide produce noise at discrete frequencies,
			which can be notch-filtered efficiently to extend coherence times of electron
			spin qubits to nearly 1ms},\ }\href {https://doi.org/10.1038/nnano.2016.170}
	{\bibfield  {journal} {\bibinfo  {journal} {Nature Nanotechnology}\ }\textbf
		{\bibinfo {volume} {12}},\ \bibinfo {pages} {16} (\bibinfo {year}
		{2017})}\BibitemShut {NoStop}%
	\bibitem [{\citenamefont {Knee}\ \emph {et~al.}(2016)\citenamefont {Knee},
		\citenamefont {Kakuyanagi}, \citenamefont {Yeh}, \citenamefont {Matsuzaki},
		\citenamefont {Toida}, \citenamefont {Yamaguchi}, \citenamefont {Saito},
		\citenamefont {Leggett},\ and\ \citenamefont {Munro}}]{knee16}%
	\BibitemOpen
	\bibfield  {author} {\bibinfo {author} {\bibfnamefont {G.~C.}\ \bibnamefont
			{Knee}}, \bibinfo {author} {\bibfnamefont {K.}~\bibnamefont {Kakuyanagi}},
		\bibinfo {author} {\bibfnamefont {M.-C.}\ \bibnamefont {Yeh}}, \bibinfo
		{author} {\bibfnamefont {Y.}~\bibnamefont {Matsuzaki}}, \bibinfo {author}
		{\bibfnamefont {H.}~\bibnamefont {Toida}}, \bibinfo {author} {\bibfnamefont
			{H.}~\bibnamefont {Yamaguchi}}, \bibinfo {author} {\bibfnamefont
			{S.}~\bibnamefont {Saito}}, \bibinfo {author} {\bibfnamefont {A.~J.}\
			\bibnamefont {Leggett}},\ and\ \bibinfo {author} {\bibfnamefont {W.~J.}\
			\bibnamefont {Munro}},\ }\bibfield  {title} {\bibinfo {title} {A strict
			experimental test of macroscopic realism in a superconducting flux qubit},\
	}\href {https://doi.org/10.1038/ncomms13253} {\bibfield  {journal} {\bibinfo
			{journal} {Nature Communications}\ }\textbf {\bibinfo {volume} {7}},\
		\bibinfo {pages} {13253} (\bibinfo {year} {2016})}\BibitemShut {NoStop}%
	\bibitem [{\citenamefont {Wu}\ \emph {et~al.}(2018)\citenamefont {Wu},
		\citenamefont {Yang}, \citenamefont {Gong}, \citenamefont {Zheng},
		\citenamefont {Deng}, \citenamefont {Yan}, \citenamefont {Zhao},
		\citenamefont {Huang}, \citenamefont {Castellano}, \citenamefont {Munro},
		\citenamefont {Nemoto}, \citenamefont {Zheng}, \citenamefont {Sun},
		\citenamefont {Liu}, \citenamefont {Zhu},\ and\ \citenamefont
		{Lu}}]{yulin18}%
	\BibitemOpen
	\bibfield  {author} {\bibinfo {author} {\bibfnamefont {Y.}~\bibnamefont
			{Wu}}, \bibinfo {author} {\bibfnamefont {L.-P.}\ \bibnamefont {Yang}},
		\bibinfo {author} {\bibfnamefont {M.}~\bibnamefont {Gong}}, \bibinfo {author}
		{\bibfnamefont {Y.}~\bibnamefont {Zheng}}, \bibinfo {author} {\bibfnamefont
			{H.}~\bibnamefont {Deng}}, \bibinfo {author} {\bibfnamefont {Z.}~\bibnamefont
			{Yan}}, \bibinfo {author} {\bibfnamefont {Y.}~\bibnamefont {Zhao}}, \bibinfo
		{author} {\bibfnamefont {K.}~\bibnamefont {Huang}}, \bibinfo {author}
		{\bibfnamefont {A.~D.}\ \bibnamefont {Castellano}}, \bibinfo {author}
		{\bibfnamefont {W.~J.}\ \bibnamefont {Munro}}, \bibinfo {author}
		{\bibfnamefont {K.}~\bibnamefont {Nemoto}}, \bibinfo {author} {\bibfnamefont
			{D.-N.}\ \bibnamefont {Zheng}}, \bibinfo {author} {\bibfnamefont {C.~P.}\
			\bibnamefont {Sun}}, \bibinfo {author} {\bibfnamefont {Y.-x.}\ \bibnamefont
			{Liu}}, \bibinfo {author} {\bibfnamefont {X.}~\bibnamefont {Zhu}},\ and\
		\bibinfo {author} {\bibfnamefont {L.}~\bibnamefont {Lu}},\ }\bibfield
	{title} {\bibinfo {title} {An efficient and compact switch for quantum
			circuits},\ }\href {https://doi.org/10.1038/s41534-018-0099-6} {\bibfield
		{journal} {\bibinfo  {journal} {npj Quantum Information}\ }\textbf {\bibinfo
			{volume} {4}},\ \bibinfo {pages} {50} (\bibinfo {year} {2018})}\BibitemShut
	{NoStop}%
	\bibitem [{\citenamefont {Touzard}\ \emph {et~al.}(2019)\citenamefont
		{Touzard}, \citenamefont {Kou}, \citenamefont {Frattini}, \citenamefont
		{Sivak}, \citenamefont {Puri}, \citenamefont {Grimm}, \citenamefont
		{Frunzio}, \citenamefont {Shankar},\ and\ \citenamefont
		{Devoret}}]{touzard19}%
	\BibitemOpen
	\bibfield  {author} {\bibinfo {author} {\bibfnamefont {S.}~\bibnamefont
			{Touzard}}, \bibinfo {author} {\bibfnamefont {A.}~\bibnamefont {Kou}},
		\bibinfo {author} {\bibfnamefont {N.~E.}\ \bibnamefont {Frattini}}, \bibinfo
		{author} {\bibfnamefont {V.~V.}\ \bibnamefont {Sivak}}, \bibinfo {author}
		{\bibfnamefont {S.}~\bibnamefont {Puri}}, \bibinfo {author} {\bibfnamefont
			{A.}~\bibnamefont {Grimm}}, \bibinfo {author} {\bibfnamefont
			{L.}~\bibnamefont {Frunzio}}, \bibinfo {author} {\bibfnamefont
			{S.}~\bibnamefont {Shankar}},\ and\ \bibinfo {author} {\bibfnamefont {M.~H.}\
			\bibnamefont {Devoret}},\ }\bibfield  {title} {\bibinfo {title} {Gated
			conditional displacement readout of superconducting qubits},\ }\href
	{https://doi.org/10.1103/PhysRevLett.122.080502} {\bibfield  {journal}
		{\bibinfo  {journal} {Phys. Rev. Lett.}\ }\textbf {\bibinfo {volume} {122}},\
		\bibinfo {pages} {080502} (\bibinfo {year} {2019})}\BibitemShut {NoStop}%
	\bibitem [{\citenamefont {Mazurek}\ \emph {et~al.}(2014)\citenamefont
		{Mazurek}, \citenamefont {Roszak}, \citenamefont {Chhajlany},\ and\
		\citenamefont {Horodecki}}]{mazurek14b}%
	\BibitemOpen
	\bibfield  {author} {\bibinfo {author} {\bibfnamefont {P.}~\bibnamefont
			{Mazurek}}, \bibinfo {author} {\bibfnamefont {K.}~\bibnamefont {Roszak}},
		\bibinfo {author} {\bibfnamefont {R.~W.}\ \bibnamefont {Chhajlany}},\ and\
		\bibinfo {author} {\bibfnamefont {P.}~\bibnamefont {Horodecki}},\ }\bibfield
	{title} {\bibinfo {title} {Sensitivity of entanglement decay of quantum-dot
			spin qubits to the external magnetic field},\ }\href
	{https://doi.org/10.1103/PhysRevA.89.062318} {\bibfield  {journal} {\bibinfo
			{journal} {Phys. Rev. A}\ }\textbf {\bibinfo {volume} {89}},\ \bibinfo
		{pages} {062318} (\bibinfo {year} {2014})}\BibitemShut {NoStop}%
	\bibitem [{\citenamefont {Zurek}(1981)}]{zurek81}%
	\BibitemOpen
	\bibfield  {author} {\bibinfo {author} {\bibfnamefont {W.~H.}\ \bibnamefont
			{Zurek}},\ }\bibfield  {title} {\bibinfo {title} {Pointer basis of quantum
			apparatus: Into what mixture does the wave packet collapse?},\ }\href
	{https://doi.org/10.1103/PhysRevD.24.1516} {\bibfield  {journal} {\bibinfo
			{journal} {Phys. Rev. D}\ }\textbf {\bibinfo {volume} {24}},\ \bibinfo
		{pages} {1516} (\bibinfo {year} {1981})}\BibitemShut {NoStop}%
	\bibitem [{\citenamefont {Zurek}(2003)}]{zurek03}%
	\BibitemOpen
	\bibfield  {author} {\bibinfo {author} {\bibfnamefont {W.~H.}\ \bibnamefont
			{Zurek}},\ }\bibfield  {title} {\bibinfo {title} {Decoherence, einselection,
			and the quantum origins of the classical},\ }\href
	{https://doi.org/10.1103/RevModPhys.75.715} {\bibfield  {journal} {\bibinfo
			{journal} {Rev. Mod. Phys.}\ }\textbf {\bibinfo {volume} {75}},\ \bibinfo
		{pages} {715} (\bibinfo {year} {2003})}\BibitemShut {NoStop}%
	\bibitem [{\citenamefont {Eisert}\ and\ \citenamefont
		{Plenio}(2002)}]{eisert02}%
	\BibitemOpen
	\bibfield  {author} {\bibinfo {author} {\bibfnamefont {J.}~\bibnamefont
			{Eisert}}\ and\ \bibinfo {author} {\bibfnamefont {M.~B.}\ \bibnamefont
			{Plenio}},\ }\bibfield  {title} {\bibinfo {title} {Quantum and classical
			correlations in quantum brownian motion},\ }\href
	{https://doi.org/10.1103/PhysRevLett.89.137902} {\bibfield  {journal}
		{\bibinfo  {journal} {Phys. Rev. Lett.}\ }\textbf {\bibinfo {volume} {89}},\
		\bibinfo {pages} {137902} (\bibinfo {year} {2002})}\BibitemShut {NoStop}%
	\bibitem [{\citenamefont {Terhal}(2000)}]{terhal00}%
	\BibitemOpen
	\bibfield  {author} {\bibinfo {author} {\bibfnamefont {B.~M.}\ \bibnamefont
			{Terhal}},\ }\bibfield  {title} {\bibinfo {title} {Bell inequalities and the
			separability criterion},\ }\href
	{https://doi.org/https://doi.org/10.1016/S0375-9601(00)00401-1} {\bibfield
		{journal} {\bibinfo  {journal} {Physics Letters A}\ }\textbf {\bibinfo
			{volume} {271}},\ \bibinfo {pages} {319} (\bibinfo {year}
		{2000})}\BibitemShut {NoStop}%
	\bibitem [{\citenamefont {Lewenstein}\ \emph {et~al.}(2000)\citenamefont
		{Lewenstein}, \citenamefont {Kraus}, \citenamefont {Cirac},\ and\
		\citenamefont {Horodecki}}]{lewenstein00}%
	\BibitemOpen
	\bibfield  {author} {\bibinfo {author} {\bibfnamefont {M.}~\bibnamefont
			{Lewenstein}}, \bibinfo {author} {\bibfnamefont {B.}~\bibnamefont {Kraus}},
		\bibinfo {author} {\bibfnamefont {J.~I.}\ \bibnamefont {Cirac}},\ and\
		\bibinfo {author} {\bibfnamefont {P.}~\bibnamefont {Horodecki}},\ }\bibfield
	{title} {\bibinfo {title} {Optimization of entanglement witnesses},\ }\href
	{https://doi.org/10.1103/PhysRevA.62.052310} {\bibfield  {journal} {\bibinfo
			{journal} {Phys. Rev. A}\ }\textbf {\bibinfo {volume} {62}},\ \bibinfo
		{pages} {052310} (\bibinfo {year} {2000})}\BibitemShut {NoStop}%
	\bibitem [{\citenamefont {G\"uhne}\ \emph {et~al.}(2002)\citenamefont
		{G\"uhne}, \citenamefont {Hyllus}, \citenamefont {Bru\ss{}}, \citenamefont
		{Ekert}, \citenamefont {Lewenstein}, \citenamefont {Macchiavello},\ and\
		\citenamefont {Sanpera}}]{guhne02}%
	\BibitemOpen
	\bibfield  {author} {\bibinfo {author} {\bibfnamefont {O.}~\bibnamefont
			{G\"uhne}}, \bibinfo {author} {\bibfnamefont {P.}~\bibnamefont {Hyllus}},
		\bibinfo {author} {\bibfnamefont {D.}~\bibnamefont {Bru\ss{}}}, \bibinfo
		{author} {\bibfnamefont {A.}~\bibnamefont {Ekert}}, \bibinfo {author}
		{\bibfnamefont {M.}~\bibnamefont {Lewenstein}}, \bibinfo {author}
		{\bibfnamefont {C.}~\bibnamefont {Macchiavello}},\ and\ \bibinfo {author}
		{\bibfnamefont {A.}~\bibnamefont {Sanpera}},\ }\bibfield  {title} {\bibinfo
		{title} {Detection of entanglement with few local measurements},\ }\href
	{https://doi.org/10.1103/PhysRevA.66.062305} {\bibfield  {journal} {\bibinfo
			{journal} {Phys. Rev. A}\ }\textbf {\bibinfo {volume} {66}},\ \bibinfo
		{pages} {062305} (\bibinfo {year} {2002})}\BibitemShut {NoStop}%
	\bibitem [{\citenamefont {Barbieri}\ \emph {et~al.}(2003)\citenamefont
		{Barbieri}, \citenamefont {De~Martini}, \citenamefont {Di~Nepi},
		\citenamefont {Mataloni}, \citenamefont {D'Ariano},\ and\ \citenamefont
		{Macchiavello}}]{barbieri03}%
	\BibitemOpen
	\bibfield  {author} {\bibinfo {author} {\bibfnamefont {M.}~\bibnamefont
			{Barbieri}}, \bibinfo {author} {\bibfnamefont {F.}~\bibnamefont
			{De~Martini}}, \bibinfo {author} {\bibfnamefont {G.}~\bibnamefont {Di~Nepi}},
		\bibinfo {author} {\bibfnamefont {P.}~\bibnamefont {Mataloni}}, \bibinfo
		{author} {\bibfnamefont {G.~M.}\ \bibnamefont {D'Ariano}},\ and\ \bibinfo
		{author} {\bibfnamefont {C.}~\bibnamefont {Macchiavello}},\ }\bibfield
	{title} {\bibinfo {title} {Detection of entanglement with polarized photons:
			Experimental realization of an entanglement witness},\ }\href
	{https://doi.org/10.1103/PhysRevLett.91.227901} {\bibfield  {journal}
		{\bibinfo  {journal} {Phys. Rev. Lett.}\ }\textbf {\bibinfo {volume} {91}},\
		\bibinfo {pages} {227901} (\bibinfo {year} {2003})}\BibitemShut {NoStop}%
	\bibitem [{\citenamefont {Doherty}\ \emph {et~al.}(2013)\citenamefont
		{Doherty}, \citenamefont {Manson}, \citenamefont {Delaney}, \citenamefont
		{Jelezko}, \citenamefont {Wrachtrup},\ and\ \citenamefont
		{Hollenberg}}]{doherty13}%
	\BibitemOpen
	\bibfield  {author} {\bibinfo {author} {\bibfnamefont {M.~W.}\ \bibnamefont
			{Doherty}}, \bibinfo {author} {\bibfnamefont {N.~B.}\ \bibnamefont {Manson}},
		\bibinfo {author} {\bibfnamefont {P.}~\bibnamefont {Delaney}}, \bibinfo
		{author} {\bibfnamefont {F.}~\bibnamefont {Jelezko}}, \bibinfo {author}
		{\bibfnamefont {J.}~\bibnamefont {Wrachtrup}},\ and\ \bibinfo {author}
		{\bibfnamefont {L.~C.}\ \bibnamefont {Hollenberg}},\ }\bibfield  {title}
	{\bibinfo {title} {The nitrogen-vacancy colour centre in diamond},\ }\href
	{https://doi.org/https://doi.org/10.1016/j.physrep.2013.02.001} {\bibfield
		{journal} {\bibinfo  {journal} {Physics Reports}\ }\textbf {\bibinfo {volume}
			{528}},\ \bibinfo {pages} {1} (\bibinfo {year} {2013})},\ \bibinfo {note}
	{the nitrogen-vacancy colour centre in diamond}\BibitemShut {NoStop}%
	\bibitem [{\citenamefont {Wood}\ \emph {et~al.}(2018)\citenamefont {Wood},
		\citenamefont {Lilette}, \citenamefont {Fein}, \citenamefont {Tomek},
		\citenamefont {McGuinness}, \citenamefont {Hollenberg}, \citenamefont
		{Scholten},\ and\ \citenamefont {Martin}}]{wood18}%
	\BibitemOpen
	\bibfield  {author} {\bibinfo {author} {\bibfnamefont {A.~A.}\ \bibnamefont
			{Wood}}, \bibinfo {author} {\bibfnamefont {E.}~\bibnamefont {Lilette}},
		\bibinfo {author} {\bibfnamefont {Y.~Y.}\ \bibnamefont {Fein}}, \bibinfo
		{author} {\bibfnamefont {N.}~\bibnamefont {Tomek}}, \bibinfo {author}
		{\bibfnamefont {L.~P.}\ \bibnamefont {McGuinness}}, \bibinfo {author}
		{\bibfnamefont {L.~C.~L.}\ \bibnamefont {Hollenberg}}, \bibinfo {author}
		{\bibfnamefont {R.~E.}\ \bibnamefont {Scholten}},\ and\ \bibinfo {author}
		{\bibfnamefont {A.~M.}\ \bibnamefont {Martin}},\ }\bibfield  {title}
	{\bibinfo {title} {Quantum measurement of a rapidly rotating spin qubit in
			diamond},\ }\bibfield  {journal} {\bibinfo  {journal} {Science Advances}\
	}\textbf {\bibinfo {volume} {4}},\ \href
	{https://doi.org/10.1126/sciadv.aar7691} {10.1126/sciadv.aar7691} (\bibinfo
	{year} {2018}),\ \Eprint
	{https://arxiv.org/abs/https://advances.sciencemag.org/content/4/5/eaar7691.full.pdf}
	{https://advances.sciencemag.org/content/4/5/eaar7691.full.pdf} \BibitemShut
	{NoStop}%
	\bibitem [{\citenamefont {Awschalom}\ \emph {et~al.}(2018)\citenamefont
		{Awschalom}, \citenamefont {Hanson}, \citenamefont {Wrachtrup},\ and\
		\citenamefont {Zhou}}]{awschalom18}%
	\BibitemOpen
	\bibfield  {author} {\bibinfo {author} {\bibfnamefont {D.~D.}\ \bibnamefont
			{Awschalom}}, \bibinfo {author} {\bibfnamefont {R.}~\bibnamefont {Hanson}},
		\bibinfo {author} {\bibfnamefont {J.}~\bibnamefont {Wrachtrup}},\ and\
		\bibinfo {author} {\bibfnamefont {B.~B.}\ \bibnamefont {Zhou}},\ }\bibfield
	{title} {\bibinfo {title} {Quantum technologies with optically interfaced
			solid-state spins},\ }\href
	{https://doi.org/https://doi.org/10.1038/s41566-018-0232-2} {\bibfield
		{journal} {\bibinfo  {journal} {Nature Photonics}\ }\textbf {\bibinfo
			{volume} {12}},\ \bibinfo {pages} {516–527} (\bibinfo {year}
		{2018})}\BibitemShut {NoStop}%
	\bibitem [{\citenamefont {Tchebotareva}\ \emph {et~al.}(2019)\citenamefont
		{Tchebotareva}, \citenamefont {Hermans}, \citenamefont {Humphreys},
		\citenamefont {Voigt}, \citenamefont {Harmsma}, \citenamefont {Cheng},
		\citenamefont {Verlaan}, \citenamefont {Dijkhuizen}, \citenamefont {de~Jong},
		\citenamefont {Dr\'eau},\ and\ \citenamefont {Hanson}}]{tchebotareva19}%
	\BibitemOpen
	\bibfield  {author} {\bibinfo {author} {\bibfnamefont {A.}~\bibnamefont
			{Tchebotareva}}, \bibinfo {author} {\bibfnamefont {S.~L.~N.}\ \bibnamefont
			{Hermans}}, \bibinfo {author} {\bibfnamefont {P.~C.}\ \bibnamefont
			{Humphreys}}, \bibinfo {author} {\bibfnamefont {D.}~\bibnamefont {Voigt}},
		\bibinfo {author} {\bibfnamefont {P.~J.}\ \bibnamefont {Harmsma}}, \bibinfo
		{author} {\bibfnamefont {L.~K.}\ \bibnamefont {Cheng}}, \bibinfo {author}
		{\bibfnamefont {A.~L.}\ \bibnamefont {Verlaan}}, \bibinfo {author}
		{\bibfnamefont {N.}~\bibnamefont {Dijkhuizen}}, \bibinfo {author}
		{\bibfnamefont {W.}~\bibnamefont {de~Jong}}, \bibinfo {author} {\bibfnamefont
			{A.}~\bibnamefont {Dr\'eau}},\ and\ \bibinfo {author} {\bibfnamefont
			{R.}~\bibnamefont {Hanson}},\ }\bibfield  {title} {\bibinfo {title}
		{Entanglement between a diamond spin qubit and a photonic time-bin qubit at
			telecom wavelength},\ }\href {https://doi.org/10.1103/PhysRevLett.123.063601}
	{\bibfield  {journal} {\bibinfo  {journal} {Phys. Rev. Lett.}\ }\textbf
		{\bibinfo {volume} {123}},\ \bibinfo {pages} {063601} (\bibinfo {year}
		{2019})}\BibitemShut {NoStop}%
	\bibitem [{\citenamefont {Fischer}\ \emph {et~al.}(2013)\citenamefont
		{Fischer}, \citenamefont {Bretschneider}, \citenamefont {London},
		\citenamefont {Budker}, \citenamefont {Gershoni},\ and\ \citenamefont
		{Frydman}}]{fischer13}%
	\BibitemOpen
	\bibfield  {author} {\bibinfo {author} {\bibfnamefont {R.}~\bibnamefont
			{Fischer}}, \bibinfo {author} {\bibfnamefont {C.~O.}\ \bibnamefont
			{Bretschneider}}, \bibinfo {author} {\bibfnamefont {P.}~\bibnamefont
			{London}}, \bibinfo {author} {\bibfnamefont {D.}~\bibnamefont {Budker}},
		\bibinfo {author} {\bibfnamefont {D.}~\bibnamefont {Gershoni}},\ and\
		\bibinfo {author} {\bibfnamefont {L.}~\bibnamefont {Frydman}},\ }\bibfield
	{title} {\bibinfo {title} {Bulk nuclear polarization enhanced at room
			temperature by optical pumping},\ }\href
	{https://doi.org/10.1103/PhysRevLett.111.057601} {\bibfield  {journal}
		{\bibinfo  {journal} {Phys. Rev. Lett.}\ }\textbf {\bibinfo {volume} {111}},\
		\bibinfo {pages} {057601} (\bibinfo {year} {2013})}\BibitemShut {NoStop}%
	\bibitem [{\citenamefont {London}\ \emph {et~al.}(2013)\citenamefont {London},
		\citenamefont {Scheuer}, \citenamefont {Cai}, \citenamefont {Schwarz},
		\citenamefont {Retzker}, \citenamefont {Plenio}, \citenamefont {Katagiri},
		\citenamefont {Teraji}, \citenamefont {Koizumi}, \citenamefont {Isoya},
		\citenamefont {Fischer}, \citenamefont {McGuinness}, \citenamefont
		{Naydenov},\ and\ \citenamefont {Jelezko}}]{london13}%
	\BibitemOpen
	\bibfield  {author} {\bibinfo {author} {\bibfnamefont {P.}~\bibnamefont
			{London}}, \bibinfo {author} {\bibfnamefont {J.}~\bibnamefont {Scheuer}},
		\bibinfo {author} {\bibfnamefont {J.-M.}\ \bibnamefont {Cai}}, \bibinfo
		{author} {\bibfnamefont {I.}~\bibnamefont {Schwarz}}, \bibinfo {author}
		{\bibfnamefont {A.}~\bibnamefont {Retzker}}, \bibinfo {author} {\bibfnamefont
			{M.~B.}\ \bibnamefont {Plenio}}, \bibinfo {author} {\bibfnamefont
			{M.}~\bibnamefont {Katagiri}}, \bibinfo {author} {\bibfnamefont
			{T.}~\bibnamefont {Teraji}}, \bibinfo {author} {\bibfnamefont
			{S.}~\bibnamefont {Koizumi}}, \bibinfo {author} {\bibfnamefont
			{J.}~\bibnamefont {Isoya}}, \bibinfo {author} {\bibfnamefont
			{R.}~\bibnamefont {Fischer}}, \bibinfo {author} {\bibfnamefont {L.~P.}\
			\bibnamefont {McGuinness}}, \bibinfo {author} {\bibfnamefont
			{B.}~\bibnamefont {Naydenov}},\ and\ \bibinfo {author} {\bibfnamefont
			{F.}~\bibnamefont {Jelezko}},\ }\bibfield  {title} {\bibinfo {title}
		{Detecting and polarizing nuclear spins with double resonance on a single
			electron spin},\ }\href {https://doi.org/10.1103/PhysRevLett.111.067601}
	{\bibfield  {journal} {\bibinfo  {journal} {Phys. Rev. Lett.}\ }\textbf
		{\bibinfo {volume} {111}},\ \bibinfo {pages} {067601} (\bibinfo {year}
		{2013})}\BibitemShut {NoStop}%
	\bibitem [{\citenamefont {Hovav}\ \emph {et~al.}(2018)\citenamefont {Hovav},
		\citenamefont {Naydenov}, \citenamefont {Jelezko},\ and\ \citenamefont
		{Bar-Gill}}]{hovav18}%
	\BibitemOpen
	\bibfield  {author} {\bibinfo {author} {\bibfnamefont {Y.}~\bibnamefont
			{Hovav}}, \bibinfo {author} {\bibfnamefont {B.}~\bibnamefont {Naydenov}},
		\bibinfo {author} {\bibfnamefont {F.}~\bibnamefont {Jelezko}},\ and\ \bibinfo
		{author} {\bibfnamefont {N.}~\bibnamefont {Bar-Gill}},\ }\bibfield  {title}
	{\bibinfo {title} {Low-field nuclear polarization using nitrogen vacancy
			centers in diamonds},\ }\href
	{https://doi.org/10.1103/PhysRevLett.120.060405} {\bibfield  {journal}
		{\bibinfo  {journal} {Phys. Rev. Lett.}\ }\textbf {\bibinfo {volume} {120}},\
		\bibinfo {pages} {060405} (\bibinfo {year} {2018})}\BibitemShut {NoStop}%
\end{thebibliography}
\end{document}